\documentclass{aa}
\usepackage{graphicx}
\usepackage{natbib}
\usepackage{graphicx}
\usepackage{CJK}
\usepackage{color}

\begin{document}
	\title{An extreme ultraviolet wave associated with the possible expansion of sheared arcades}
	\author{Yihan Liu\inst{1}, Ruisheng Zheng\inst{1,2}, Liang Zhang\inst{1}, Hengyuan Wei\inst{1}, Ze Zhong\inst{2}, Shuhong Yang\inst{3,4}, and Yao Chen\inst{1,2}}
	
	\institute{Institute of Space Sciences, Shandong University, Weihai 264209, China; \\
		\email{ruishengzheng@sdu.edu.cn}(Ruisheng Zheng)\\
		\and
		Institute of Frontier and Interdisciplinary Science, Shandong University, Qingdao 266237, China; \\
		\and
		CAS Key Laboratory of Solar Activity, National Astronomical Observatories, Chinese Academy of Sciences, Beijing 100101, China;\\
		\and
		School of Astronomy and Space Science, University of Chinese Academy of Sciences, Beijing 100049, China;
	}
	
	\date{Received / Accepted }
	\titlerunning{EUV Wave associated with expansion of sheared arcades}
	\authorrunning{Liu et al.}
	
	\abstract
	{
		Solar extreme ultraviolet (EUV) waves are propagating disturbances in the corona, and they usually accompany with various solar eruptions, from large-scale coronal mass ejections to small-scale coronal jets.
	}
	{
		Generally, it is believed that EUV waves are driven by the rapid expansion of coronal loops overlying the erupting cores. In this Letter, we present an exception of EUV wave that was not triggered by the expansion of coronal loops overlying the erupting core. 
	}
	{ 
		Combining the multiwavelength observations from multiple instruments, we studied the event in detail.
	}
	{The eruption was restricted in the active region (AR) and disturbed the nearby sheared arcades (SAs) connecting the source AR to a remote AR. Interestingly, following the disturbance, an EUV wave formed close to the SAs, but far away from the eruption source. 
	}
	{All the results showed that the EUV wave had a closer temporal and spatial relationship with the disappearing part of SAs than the confined eruption. Hence, we suggest that the EUV wave was likely triggered by the expansion of some strands of SAs, rather than the expansion of erupting loops. It can be a possible complement for the driving mechanisms of EUV waves.
	}

	
	\maketitle
	\section{Introduction}
	Solar extreme ultraviolet (EUV) waves are spectacular disturbance in the corona, and propagate with speeds in the range of 200--1500 km s$^{-1}$ \citep{2013ApJ...776...58N}. Their wavefronts appear bright in 193 {\bf\rm\AA} (Fe XII and Fe XXIV) and 211 {\bf\rm\AA} (Fe XIV) passbands, and sometimes become dark in 171 {\bf\rm\AA}  (Fe IX) passband \citep{2015LRSP...12....3W}. EUV waves are closely integrated with diverse solar activities, such as filament eruptions, coronal mass ejections (CMEs), and coronal jets \citep{2019ApJ...871..232Z}, and can be links between sympathetic activities in different active regions (ARs) \citep{2019ApJ...883...32C}. Furthermore, the study of EUV waves is important for the understanding of corona heating and particle accelerations, and the application in coronal seismology\citep{2014SoPh..289.3233L}. 
	
	As observational evidences and numerical experiments accumulated, it is generally believed that an EUV wave consists of a leading fast-mode magneto-sonic wave (or shock) component and a following non-wave CME component \citep{2014SoPh..289.3233L}. The fast-mode components are usually driven by rapid swelling of CME bubbles \citep{2020ChSBu..65.3909S}, and the rapid swelling of CME bubbles can occur during the events of solar flares, filament eruptions \citep{2017ApJ...851..101S}, coronal jets \citep{2022ApJ...931..162Z}, or even untwisting motion of filament helical structures \citep{2019ApJ...873...22S}. For the confined eruptions \citep{2012ApJ...753..112Z, 2013ApJ...778...70C, 2018ApJ...860L...8S, 2019ApJ...871..232Z, 2020ApJ...894..139Z}, EUV waves can be still triggered by expansions of coronal loops overlying erupting cores, even if there was no CME. Hence, the expansion of coronal loops overlying erupting cores is the direct key for the formation of EUV waves.

 { In this Letter, we present an exception of EUV wave generated after a confined eruption on 2015 December 11. Observations show the wave had a closer temporal and spatial relationship with the disappearing sheared arcades (SAs) than the eruption, imply the wave is more likely triggered by the SAs, which disturbed by the confined eruption.}
 

    \section{Data and Method}
    \subsection{Data}
    { The EUV observations for this event are mainly employed from Atmospheric Imaging Assembly (AIA; \cite{2012SoPh..275....3P}) onboard the {\it Solar Dynamic Observatory (SDO)} spacecraft. Each AIA image has a cadence of 12 second and spatial resolution of 1.2$^{\prime\prime} $ (0.6$^{\prime\prime} $ per pixel), which provide full disc solar image. The filaments are also confirmed by the H$\rm\alpha$ data from the New Vacuum Solar Telescope (NVST; \cite{2014RAA....14..705L}) in {\it Fuxian Solar Observatory of Yunnan Astronomical Observatory, CAS}. The H$\rm\alpha$ images have a pixel size of 0.165$^{\prime\prime} $, and a 12 s temporal resolution, and the field of view (FOV) is 150$^{\prime\prime} $. 

    The magnetic field configuration of the eruption source is checked by the observations from Helioseismic and Magnetic Imager (HMI; \cite{2012SoPh..275..207S}) onboard SDO. HMI provides vector magnetic field with a cadence of 720 second and spatial resolution of 1 $^{\prime\prime} $ (0.5$^{\prime\prime}$ per pixel) of the full solar disc. In addition, the associated CME was detected by Large Angle and Spectrometric Coronagraph (LASCO, \cite{1995SoPh..162..357B}) onboard the {\it Solar and Heliospheric Observatory (SOHO)} spacecraft, and the related flares are recorded by the Geostationary Operational Environment Satellite (GOES) in the form of integrated full-disk soft X-ray emission from the Sun. We also used data in hard X-ray channel from the Reuven Ramaty High Energy Solar Spectroscopic Imager (RHESSI; \cite{2002SoPh..210....3L}) to shoe the energy release during the flare.}

    \subsection{Method}
    {For the AIA EUV observations, we used preparing program from aiapy\citep{Barnes2020}, including the step of deconvolution with the point spread function (PSF). To clarify the sheared arcades, we used multi-scale Gauss normalization (MGN; \cite{2014SoPh..289.2945M}). All images were drawn via sunpy\citep{sunpy_community2020} and astropy\citep{2013A&A...558A..33A,2018AJ....156..123A}.

    The data from RHESSI were reconstructed by VIS\_CS algorithm \citep{2017ApJ...849...10F}. To get vector magnetic field of the solar corona, we used non-linear force-free field (NLFFF; \cite{2021LRSP...18....1W}) extrapolation with the HMI vector field data as the boundary condition, with the code of Yang Guo
\footnote{\url{https://github.com/njuguoyang/magnetic_modeling_codes}} \citep{2017ScChD..60.1408G}. To choose suitable magnetic field lines, we used the code of Rui Liu to calculate Q-factor and twist number\footnote{\url{http://staff.ustc.edu.cn/~rliu/qfactor.html}}
    \citep{2016ApJ...818..148L}. The 3D field lines were drawn by pyvista \citep{sullivan2019pyvista}.}

	\section{Results and Analysis}
	\subsection{Confined Eruption}  \label{section:Confined Eruption}
	Figure \ref{fig:1} shows the pre-eruption environment in the { AIA} and { NVST}. In AIA 171 {\bf\rm\AA} (panel (a)), it is clear for the existence of the magnetic SAs (the purple dashed curve), connecting the eruption source of { National Oceanic and Atmospheric Administration (NOAA)} AR 12465 (the white square) and the remote NOAA AR 12464 (the cyan square). The invisibility of the SAs in AIA 94 {\bf\rm\AA} (Fe XVIII, panel (b)) indicates their low-temperature of $\sim$0.6 MK. Using the NLFFF extrapolation method, the configuration of NOAA AR 12465 and nearby SAs (green) were reconstructed and superimposed on the HMI vector magnetogram at 16:36 UT (panel (c)). Moreover, two filaments (red and pink) involving with the subsequent eruption were found underlying the AR loops (blue), and NOAA AR 12465 is also shown in an enlarged view in the top-right corner of panel (c). Hours before the eruption, the highly-twisted filament (F1) in the AR center (blue arrows) and another filament (F2) in the AR boundary (green arrows) were clearly distinguished in AIA 304 {\bf\rm\AA} (He II line) and H$\rm\alpha$ (panels (d)-(e)).  Interestingly, the following filament eruption was accompanied with a C9.6-class flare with double peaks. In the curves of GOES soft X-ray flux (panel (f)), the double peaks are separately at 16:53 and 17:15 UT, indicating two episodes of energy release process during the following confined eruption.
	
	The two-step energy release in the confined eruption is clearly shown in AIA 304, 94, 171 {\bf\rm\AA} in Figure \ref{fig:2} and Animation 1. In the first step (panels (a)-(c)), F1 (the white arrow) erupted and induced with compact flare ribbons (black arrows) in the AR center. In the AIA  94 {\bf\rm\AA} image overlaid with contours of { RHESSI}, { the X-ray source at the channel of 3-6 keV (orange)} were clearly over the looptop of the post-flare arcade (panel (b)). A few minutes later, the second step began (panels (d)-(f)). The erupting structure of F1 seemed to be broken (green arrows), which was likely due to the interaction with the nearby overlying loops of F2 (pink arrows). Consequently (panel (h)), a new longer flux rope (the yellow arrow) formed and began to erupt northwestwards as a result of the confinement of the overlying AR loops, and its eruption caused one more group of post-flare arcade (red arrows). On the other hand, the eruption of the longer flux rope also encountered the nearby SAs (blue arrows in right column), and directly made the SAs disappear (the right column). { Before the disappearance of SAs, some brightenings appeared around the eastern end (red arrow in panel (g)), which possibly resulted from magnetic reconnection between the SAs and the expanding structure of the newly-formed longer flux rope.} Overlaid with the contours of RHESSI, a new X-ray source of 3-6 (orange) keV was obviously above the looptop of the new group of post-flare arcade (panel (h)). Comparing with the double-peaked flare in GOES soft X-ray flux (Figure \ref{fig:1}(f)), the eruption of F1 contributed to the first flare peak, and the later eruption of newly-formed longer flux rope was associated with the second flare peak. 
	
	\subsection{Disturbance of SAs} \label{section: magnetic arc}
	
	The interaction between the erupting flux rope and the SAs (white arrows) is clearly seen in AIA images with { MGN} (Figure \ref{fig:3} and Animation 2). Before the second step of the eruption (panel (a)), the coronal loops (blue arrows) overlying the flux rope were close to the top of the SAs. During the eruption, the overlying loops expanded and interacted with the SAs (panel (b)). As a result, some strands of the SAs released and moved southwards (panel (c)), and the faint motion can be clearly seen in Animation 2. After the interaction, the expanding flux rope was finally confined (the green arrow in panel (d)), and the SAs was nearly invisible (Figure \ref{fig:2}(i). Two slices of S1 and S2 (the red and blue line in panel (d) and (c)) were chosen to analyze the evolution of the second-step eruption and the disturbance of the SAs. In the time-distance plot of running-difference AIA 171 {\bf\rm\AA} images along the S1 (panel (e)),  the overlying loops of the flux rope first expanded at a speed of $\sim$182 km s$^{-1}$ and then vanished in a few minutes, possibly because they were heated to a high temperature. In the time-distance plot in running-difference AIA 131 {\bf\rm\AA} images along the S1 (panel (f)), the flux rope became visible at the late phase of the eruption, and propagated at a low speed of $\sim$63 km s$^{-1}$ due to the strong confinement of the AR loops. In the time-distance plot in original AIA 171 {\bf\rm\AA} images along the S2 (panel (i)), the SAs showed a damped oscillation and slowly disappeared (the white arrow), and the faint strands (the red arrow) released from the SAs with a high speed of $\sim$196 km s$^{-1}$. It is likely as a result of the interaction between the erupting loops and the SAs.
	
	\subsection{EUV Wave} \label{EUV wave}
	
    Intriguingly, after the interaction between the eruption and the SAs, an EUV wave formed, best seen in difference images of AIA 171, 193, and 211 {\bf\rm\AA} (Figure \ref{fig:4} and Animation 3). Disturbed by the second step of the eruption, the SAs (the red box in panel (a)) became faint, and a newly-formed loop (black arrows in panels (b)-(c)) appeared and expanded southwestwards from the SAs, consistent with the faint threads of SAs in Figure \ref{fig:3}. In a few minutes (panels (d)-(f)), the EUV wave appeared in the southwest far away from AR 12465 (green arrows), and the SAs were replaced by dimmings (the red box). Next, the wave evolution is shown in the time-distance plots (the bottom panels) of difference images in AIA 171, 193, and 211 {\bf\rm\AA} along the wave propagation direction (S3; the blue sector in panel (e)). The wave was hardly distinguished in AIA 171 {\bf\rm\AA}, but showed a distinct bright front in AIA 193 and 211 {\bf\rm\AA} (panels (g)-(i)). The wave had a nearly constant speed of $\sim$330 km s$^{-1}$ (red dotted lines), and started close to the SAs (the region between green dashed lines indicating the boundaries of the red boxes in panels (a) and (d)) at $\sim$17:15 UT. In panel (g), the long-duration dimmings (the white arrow) reveals the disappearance of the SAs after the disturbance. In 193 and 211 {\bf\rm\AA}, the faint brightenings (cyan arrows) indicate the disturbance of the SAs. Note that there is loop-like dimmings (blue arrows) between the SAs and the wavefront, which is likely due to the expansion of the SAs. The normalized intensities (blue curves) in the region of the SAs (the red box) is overlaid on the time-distance plots. The intensities in three channels first all reached a peak at $\sim$17:06 UT and then sharply dropped. The peak time was consistent with the disturbance of the SAs, and the rapid decrease is as the result of the disappearance of the SAs. Therefore, it is clear that the wave onset has a close spatial and temporal relationship with the disturbance and expansion of the SAs. 
	
	\subsection{Associated CME} \label{CME}
    { About three hours later than the EUV wave onset, a poor CME was detected by the C2 of LASCO (Figure 5). The CME appeared as a puff-like structure, with a faint loop (white arrows) front and no erupting core (panel (a)). The puff-like CME had a linear speed\footnote{\url{https://cdaw.gsfc.nasa.gov/CME_list/UNIVERSAL/2015_12/yht/20151211.182404.w065n.v0232.p196g.yht}}
    of $\sim$232 km s$^{-1}$, and the liftoff time of the CME was traced back at $\sim$17:00 UT (panel (b)). The close temporal and spatial relationship indicates that the CME was associated with the event we studied here, while there was no other eruption at the same time.
 
    However, the eruption was confined by the overlying magnetic fields, and there was no erupting core in the puff-like CME. On the other hand, the CME liftoff was also close to the beginning time of the expansion of the SAs (the peak time of intensity curves in Figure 4). Hence, it is impossible that the CME was resulted from the confined eruption, and it is reasonable that the expanding SAs can evolved into a CME.}
	
	\section{Conclusion and Discussion}
	
\begin{figure*}
		\centering
		\includegraphics[width=12cm] {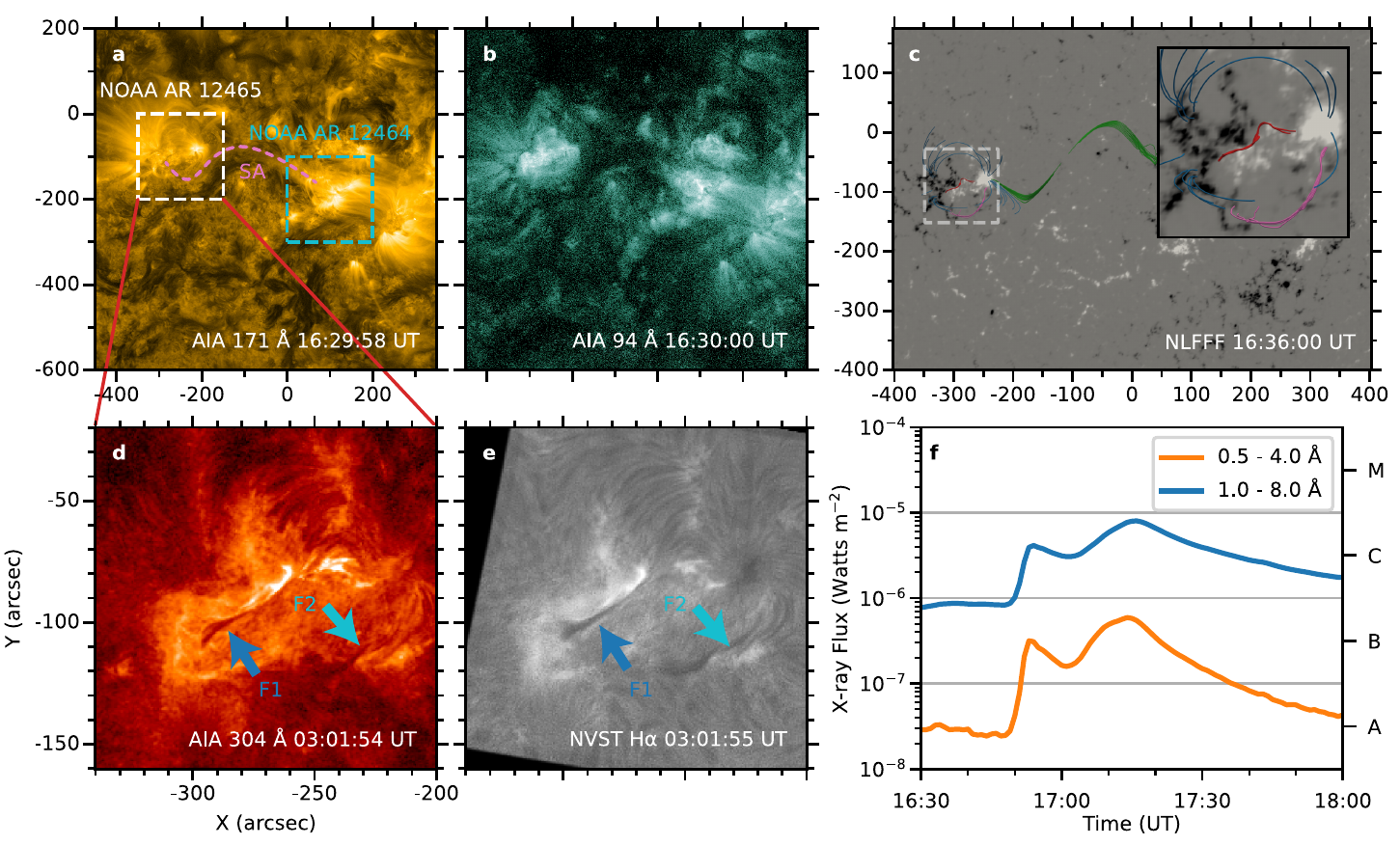}
		\caption{Overview of pre-eruption environment.
			(a)-(b) Coronal environment of NOAA ARs of 12465 (the white square) and 12464 (the cyan square) in AIA 171 and 94 {\bf\rm\AA.} The purple curve indicates the magnetic SAs. (c) NLFFF extrapolated field lines, superposing on the HMI vector magnetogram at 16:36 UT, show the filaments (red and pink), the overlying coronal loops (blue), and the SAs (green). The configuration of NOAA AR 12465 is in the enlarged view in the top-right corner. (d)-(e) The filaments of F1 (blue arrows) and F2 (green arrows) in AIA 304 {\bf\rm\AA} and NVST H$\rm\alpha$. (f) GOES soft X-ray flux of the C9.6-class flare in channel 1.0 - 8.0 {\bf\rm\AA} (the blue curve) and 0.5 - 4.0 {\bf\rm\AA} (the orange curve).  }
		\label{fig:1}
	\end{figure*}

\begin{figure*}
		\centering
		\includegraphics{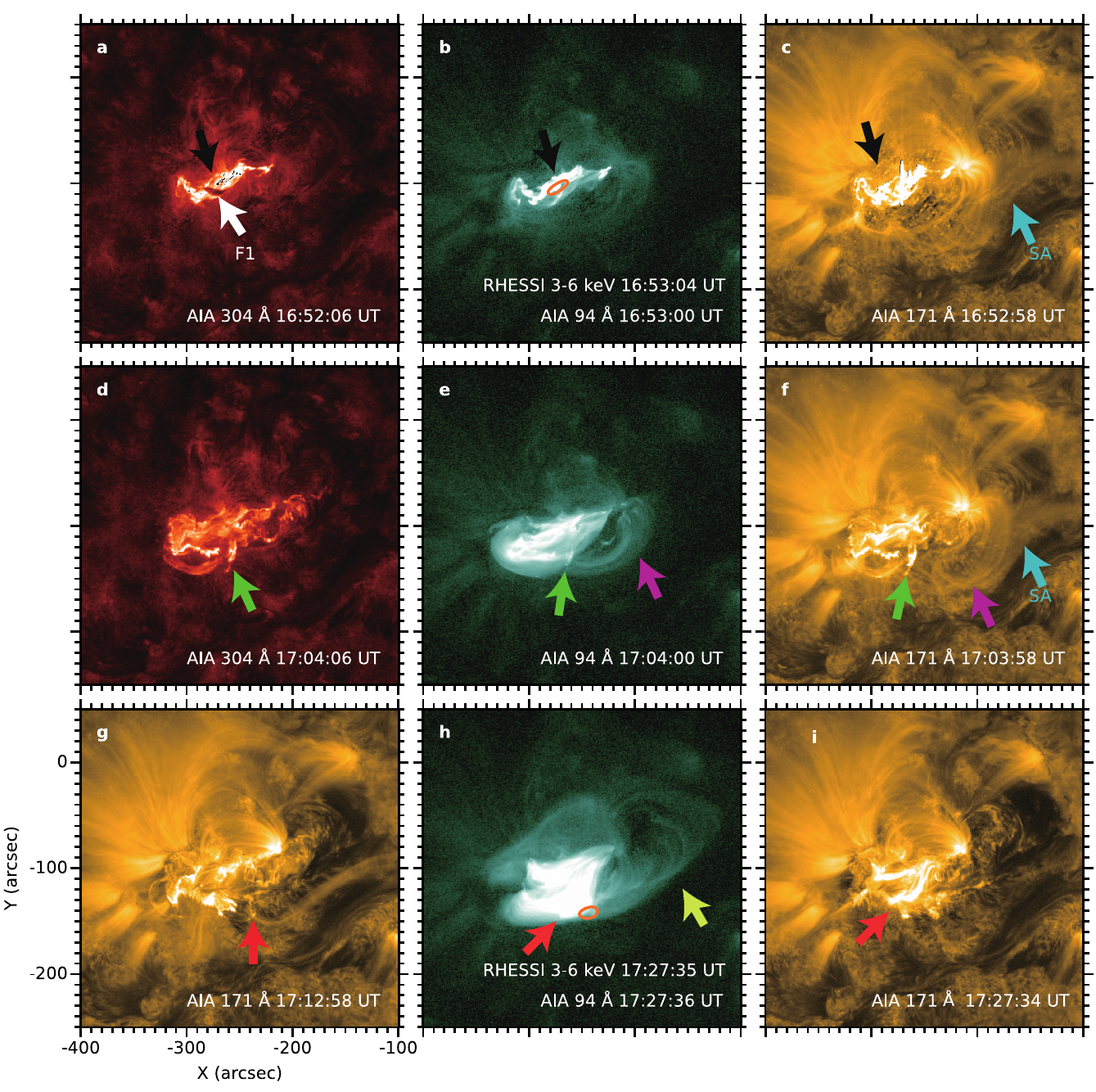}
		\caption{Two-step energy release. 
			(a)-(c) The first step. Black arrows indicate the flare ribbons, and the white and blue arrows indicate the filament of F1 and SAs, separately. The orange contour represents the X-ray source of RHESSI 3-6 keV.
			(d)-(i) The second step. { Green arrows indicate the break of the erupting structures of F1}, and pink arrows indicate the overlying coronal loops. The yellow arrow indicates the newly-formed longer flux rope, and red arrows indicate the post-flare arcade in the second step. The orange contour indicates the X-ray source of RHESSI 3-6 keV. { The red arrow indicate the brightenings that possibly related to magnetic reconnection between SAs and the longer flux rope.}}
		\label{fig:2}
	\end{figure*}

\begin{figure*}
		\centering
		\includegraphics{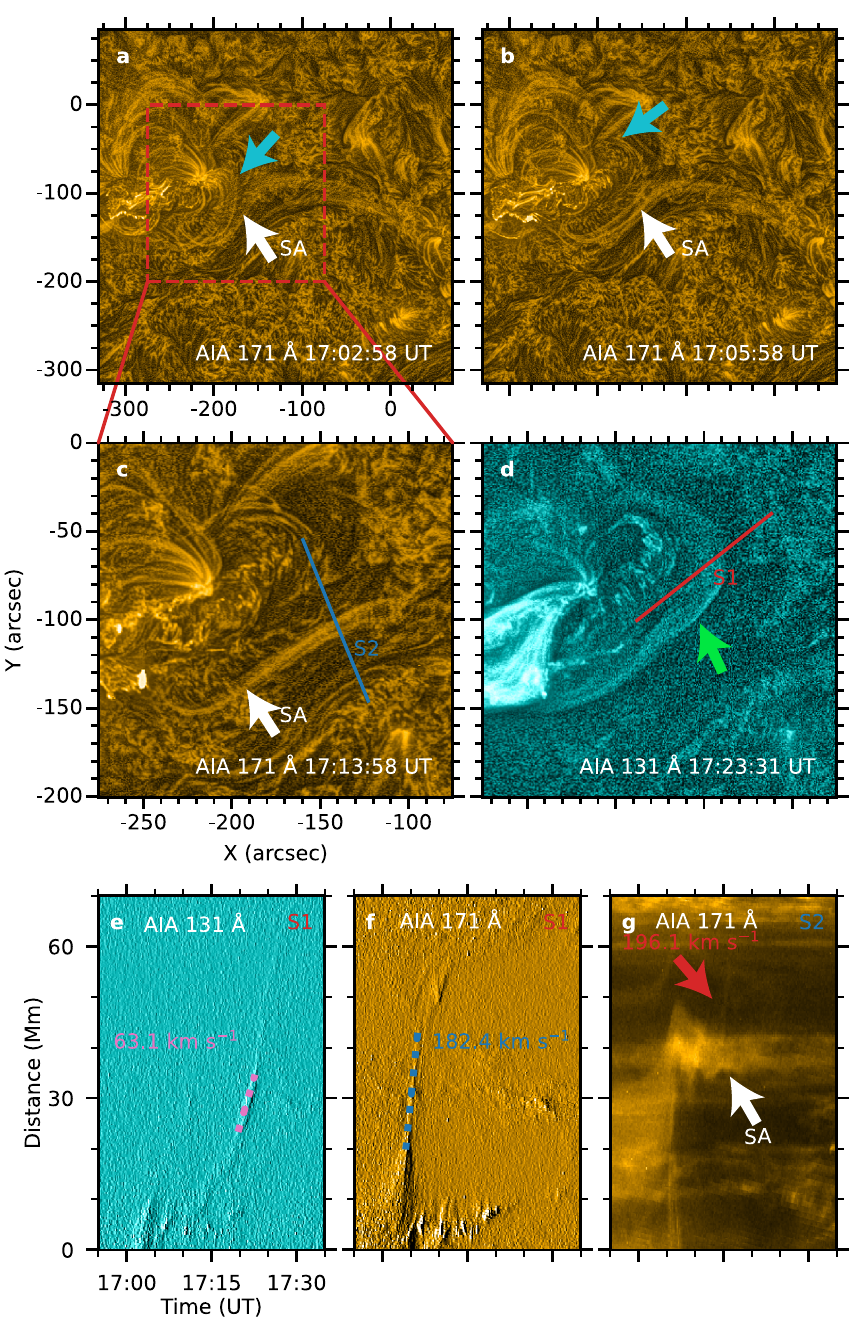}
		\caption{Interaction between the eruption and SAs. 
			(a)-(d) Interaction process. White arrows indicate SAs, and cyan arrows indicate the erupting loops wrapping the flux rope (the green arrow). The red box represents the field of view of panels (c)-(d). 
			(e)-(g) Time-distance plots along S1 and S2 (red and blue lines in panels (c)-(d)). The dotted lines show the expansion speeds of the flux rope (pink), the loops (blue), and the strands (red) of SAs (the white arrow).
		}
		\label{fig:3}
	\end{figure*}

\begin{figure*}
		\centering
		\includegraphics{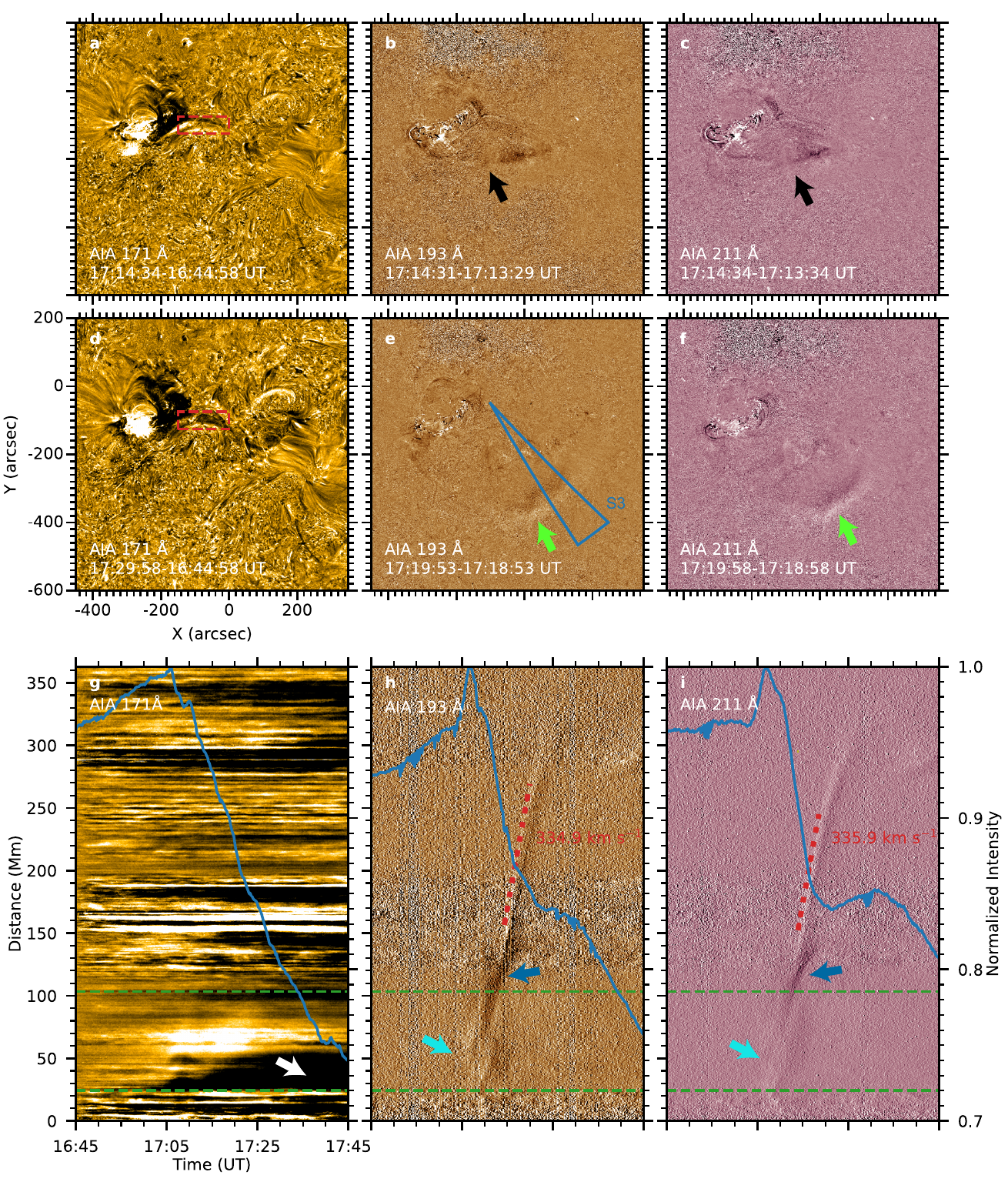}
		\caption{EUV wave in difference images of AIA 171, 193, and 211 {\bf\rm\AA}.
			(a)-(f) The red boxes show the region of SAs, and the green and black arrows separately indicate the EUV wave and the expanding loops. (g)-(i) Time-distance plots along S3 (the blue sector in panel (e)), overlaid with the normalized intensity changes (blue curves) within the red square in panels (a) and (d). The red dotted lines show the speeds of the wavefront. The white arrow indicates the long-duration dimmings, and cyan arrows indicate the faint brightening, and blue arrows indicate the loop-like dimmings. Green dashed lines represent the boundaries of the region of red square in panels (a) and (d).
		}
		\label{fig:4}
	\end{figure*}

\begin{figure*}
		\centering
		\includegraphics{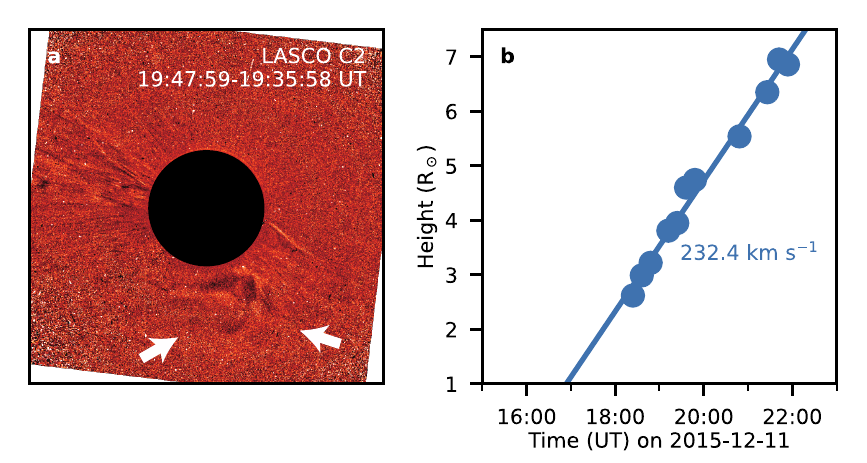}
		\caption{Following CME process. (a) CME front observed by LASCO. White arrows indicate the CME front. (b) The curve fit results of CME front}
		\label{fig:5}
	\end{figure*}

	In this Letter, we present a particular case of EUV wave that formed far away from a confined eruption. The confined eruption involved with two filaments in AR 12465 and the SAs connecting two neighboring ARs of 12465 and 12464, and was accompanied with a double-peaked C5.6-class flare (Figure \ref{fig:1}). The eruption consisted of two steps of energy release process, the earlier eruption of the twisted filament (F1) in the AR center and the latter eruption of a longer flux rope over the filament (F2) at the AR boundary (Figure \ref{fig:2})). Due to the confinement of the restricting AR fields, the lateral expansion of the longer flux rope interacted with the nearby SAs. Consequently, some weak strands released and expanded from the SAs that finally disappeared (Figure \ref{fig:3})). Intriguingly, an EUV wave formed ahead of faint expanding loops that possibly represented the releasing strands of SAs, but was far away from the eruption source (Figure \ref{fig:4}). { Finally, a detectable puff-like CME was possibly associated with the expansion of SAs (Figure \ref{fig:5}).}
	
	Generally, the formation of EUV waves are closely associated with the rapid expansion of coronal loops over the erupting cores \citep{2014SoPh..289.3233L}. However, for the case in this Letter, the two-step eruption was totally restricted in the AR cage, and the initial site of the EUV wave was $\sim$100 Mm away from the edge of the eruption source (Figure \ref{fig:4}(h)-(i)), considering that the starting site of S3 (Figure \ref{fig:4}(e)) was the expansion terminal of the confined eruption. It is difficult for the confined eruption to trigger an EUV wave at a remote region. Even if the EUV wave could be triggered by the expansion of the confined structure, it is impossible for the wave signature to first keep be faint/invisible and then became bright after the propagation of a long distance. Hence, the poor spatial and temporal relationship excludes the possibility of the confined eruption for the wave initiation.
	
	On the contrary, the wavefront was near the SAs (the region between dashed green lines in Figure \ref{fig:4}(g)-(i)), and the wave onset ($\sim$17:14 UT) was close to the time of the brightenings that were resulted from the disturbance of the confined eruption in the SAs (cyan arrows in Figure \ref{fig:4}(h)-(i)). Note that the loop-like dimmings (blue arrows in Figure \ref{fig:4}(g)-(i)) connecting the wavefront and the SAs likely indicate the expansion of the strands of SAs (black arrows in Figure \ref{fig:4}(b)-(c)), consistent with the propagation of the weak strands (the red arrow in Figure \ref{fig:3}(g)). Therefore, the EUV wave was likely driven by the quick expansion of the strands of SAs, because of its intimate spacial and temporal relationship with the SAs. 
	
	{ The interpretation of the observations given here imply a role for SAs in the generation of the EUV wave. After the interaction with the magnetic structure of the confined eruption, the SAs quickly expanded and invoked the formation of the EUV wave. The quick expansion was due to the original shear of the SAs, not induced by a erupting core (piston) as the normal way for CME-driven or jet-driven waves\citep{2015LRSP...12....3W}. Furthermore, the detection of the associated puff-like CME (Figure \ref{fig:5}(f)) can also provide the evidence of the continuous expansion of SAs that finally evolved into the faint front.}
	
\begin{figure*}[t]
		\centering
		\includegraphics{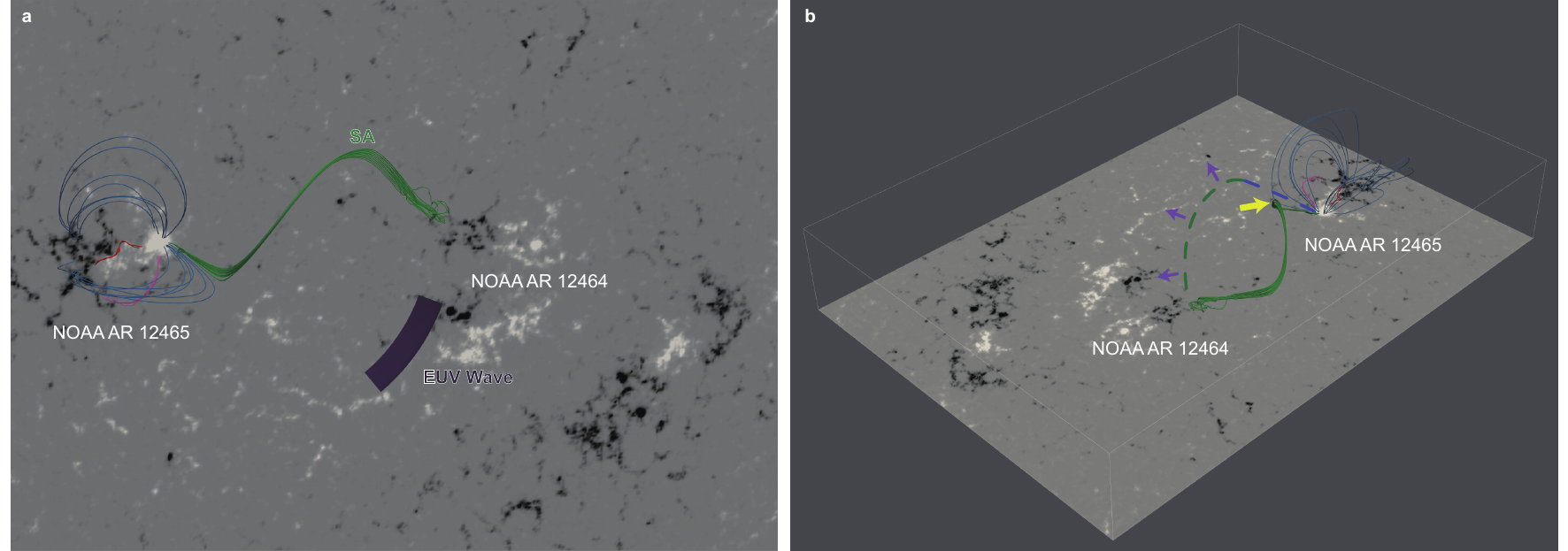}
		\caption{Schematic drawing of the possible scenario for the wave formation. (a) The relationship between the EUV wave (the deep-blue shade) and the SAs, overlaid on the HMI vector magnetogram at $\sim$16:36 UT. The magnetic structures in the source AR are as same as that in Figure \ref{fig:1} (c). (b) The possible scenario for the expansion (purple arrows) of the releasing strands of SAs (the green-blue dashed line) in a side view. The yellow arrow points out the possible sites of magnetic reconnection. An animate shows the 3D structure of NLFFF is available.}
		\label{fig:6}
	\end{figure*}

	The schematic drawing in Figure \ref{fig:6} shows the possible scenario for the wave formation. The wavefront and the SAs extrapolated from the NLFFF results are overlaid on the HMI vector magnetogram at $\sim$16:36 UT (Figure \ref{fig:6}(a)). It is obvious that the EUV wave (the deep-blue shade) is much close to the SAs (green lines), and is far away from the coronal structures in AR 12465. To explain the reason for the release of the strands of the SAs, the SAs and coronal loops in AR 12465 from the NLFFF results are displayed in a side view (Figure \ref{fig:6}(b) and Animation 4). It is possible that magnetic reconnection (the yellow arrow) between the SAs and the expanding coronal loops, following the lateral expansion of the low-lying flux rope. As a result, some strands (the green-blue dashed line) release from the SAs and expand outwards (purple arrows), and therefore the EUV wave forms ahead of the expanding strands of SAs.
	
	EUV waves in previous reports are all associated with the rapid loop expansions that are driven by the inner erupting cores. Here, the EUV wave was much less related with the loop expansion of the eruption, but was likely triggered by the expansion of SA strands that was resulted from the interaction with the nearby confined eruption. To our knowledge, it is absolutely an exception for the unusual formation manner of the EUV wave, though the magnetic reconnection contributing in the releasing of SA strands missed, due to the resolution limitation of the observations. { Numerical experiments and more observational examples are required to understand the general viability of this trigger mechanism as an explanation for at least some EUV waves.}
	
	\begin{acknowledgements}
		SDO and SOHO are missions of NASA's Living With a Star Program. The authors thank the teams of SDO, SOHO, NVST and RHESSI for providing the data. This work is supported by grants NSFC 11790303, 11973031 and 12073016.
	\end{acknowledgements}
	
	\bibliographystyle{aa} 
	\bibliography{sample631.bib} 
	
	\newpage
    \newpage
    \newpage
    \newpage
    \newpage
    \newpage
\end{document}